\documentclass[pre,showpacs,notitlepage]{revtex4}
\usepackage{graphicx}
\usepackage{dcolumn}
\usepackage{bm}
\usepackage{natbib}
\usepackage{color}

\begin{document}


\title[]{Dynamic Monte Carlo Simulations of Anisotropic Colloids}

\author{Sara Jabbari-Farouji}
\author{Emmanuel Trizac}%
\affiliation{
LPTMS, CNRS and Universit$\acute{e}$ Paris-Sud, UMR8626, Bat. 100, 91405 Orsay, France
}%


\date{\today}

\begin{abstract}
We put forward a simple procedure for extracting dynamical information from Monte Carlo
simulations, by appropriate matching of the short-time diffusion tensor
with its infinite-dilution
limit counterpart, which is supposed to be known.
This approach --discarding hydrodynamics interactions--
first allows us to improve the efficiency of previous
Dynamic Monte Carlo algorithms for spherical Brownian particles. In a second step, we address the case
of anisotropic colloids with orientational degrees of freedom. As an illustration, we present
a detailed study of the dynamics of thin platelets, with emphasis on long-time diffusion and
orientational correlations.
\end{abstract}

\keywords{Dynamic Monte Carlo, anisotropic colloids, hard disks(platelets) }
\maketitle

\bigskip

\section{\label{sec:level1} Introduction}

Monte Carlo simulations (MC) provide a powerful method for calculating the thermodynamical averages of physical quantities of many-body systems  and have been employed to study the equilibrium properties and phases of a large variety of physical systems. Such approaches rely on the generation of a Markov chain --a stochastic set of configurations-- with appropriate sampling of phase space.
In the most common variants of MC, such as Metropolis algorithms, trial moves are accepted  with a certain probability that satisfies detailed balance with respect to the desired Boltzmann distribution \cite{Allen,Frenkel}. This method is frequently used to obtain average values of  macroscopic quantities in equilibrium thermodynamic ensembles: MC methods were initially devised to study static
properties. In some case though, where mesoscopic degrees of freedom interact
with microscopic ones, the ensuing separation of time scales allows to
replace the microscopic details by a noise, in the spirit of a Langevin equation.
A suitably chosen MC scheme --stochastic in nature-- then allows to study dynamical
features of the mesoscopic degrees of freedom, meaning that correlations between successive configurations in the Markov chain of the MC simulation can be interpreted  in terms of the dynamic correlation functions. The legitimacy of this  approach  stems from the coincidence of the Fokker-Planck equations
of the original system governed by the aforementioned Langevin-type dynamics, and of the fictitious MC dynamics \cite{Weinberg,Cicho,Kikuchi}.

  Recently, there has been a rise of interest to employ MC simulations to study the dynamics of colloidal suspensions \cite{Cicho,Eduardo1,Eduardo2}. On the time-scale that momenta correlations have decayed, colloids undergo diffusive motion as a result of collisions with solvent molecules. Therefore, at these time-scales   the stochastic dynamics generated by MC algorithm seems to be more appropriate compared to the  deterministic Newtonian dynamics where solvent is omitted.  Dynamic (sometimes known as Brownian) Monte Carlo (DMC) algorithms
  in which only single particle moves with sufficiently  small displacements  are allowed,
  reproduce the real dynamics for times larger than the time-scale of momenta relaxation.  In such a case, DMC  becomes equivalent to  Brownian dynamics (BD) simulations \cite{Weinberg,Rossky,Cicho,Kikuchi,Scharetl,Heyes,Babu,BMC-emulsion,Eduardo1,Eduardo2}: a BD algorithm is also stochastic, with integrated out momenta and positions evolving with overdamped Langevin dynamics \cite{Allen}. The advantage of  DMC  over BD is that it is easily adaptable to systems with non-differentiable (hard) potentials. Although an "event-driven" variant of BD technique has been developed \cite{ED} to deal with such types of interactions, it is computationally more  cumbersome and expensive. Hence, studying the dynamics of hard particles with the DMC scheme  seems to be an efficient route, provided that an accurate mapping between the Monte Carlo time-step and the physical time is
  worked out. Achieving this goal is the main motivation of the present work.

The significance of establishing the matching of time-scales is justified by the  recent increasing use of DMC for studying dynamics of various systems \cite{Berthier,Belli,Patti,Coslovich}. Recently, it has been proposed that  equating the square of amplitude of MC displacement scaled with  acceptance probability  with infinite-dilution limit diffusion coefficient  provides a good estimate of  the physical time for spherical particles \cite{Eduardo1}.  The applicability of this proposition  is justified through  agreement of   BD simulations results with those of DMC  \cite{Eduardo1,Eduardo2}.  Furthermore,  these studies  show that  scaling  the Monte Carlo time step with acceptance probability allows one to extend the limit of validity of DMC  to  relatively larger displacement amplitudes  corresponding to  acceptance probabilities --the fraction of accepted MC attempted moves--
significantly smaller than 1.  Here, we propose  an alternative  physically motivated approach for mapping the MC time to physical time. It allows us to push the limit of applicability of DMC  to  even larger  displacement amplitudes (smaller acceptance probabilities). Our scheme is based on  equating the short-time self-diffusion  extracted from simulations  directly with  the infinite-dilution  diffusion coefficient,
Particular attention is paid to anisotropic particles. At variance with previous
approaches that did not consider the anisotropy of the short time diffusion tensor \cite{Eduardo2},
we have taken into account the important coupling between orientational and translational
degrees of freedom. In all what follows, the various short-time diffusion constants are supposed to be known, and implicitly account for the
presence of an underlying solvent.




The rest of the paper is organized as follows.
In section \ref{sec:method}, the method is presented, for both spherical and anisotropic particles.
In particular, we discuss the relation between the amplitudes of  translational and rotational moves,
essential to achieve  a physically consistent diffusive process.
In section \ref{sec:assess}, we study the convergence and self-consistency
of  DMC simulations as a
function of displacement amplitude for both spherical and disk-shaped particles, and compare our
approach with previous investigations. As an illustration, the method is employed
in section \ref{sec:anomalous} to explore the development of long-time translational diffusion  of
infinitely-thin hard  disks (platelets) as a function of density. We find that  upon increasing the
density deep in the nematic phase, the long-time diffusion becomes anisotropic, and that
in contrast to an impeded diffusion in the nematic direction, the transverse diffusion of disks is enhanced. Concluding remarks close the paper with section \ref{sec:concl}.

\section{Methodology }
\label{sec:method}

We start by describing the dynamic Monte Carlo algorithms used  for both spherical and  anisotropic particles. In each case, we  discuss  the  procedure for matching of the time scales. We then introduce the model systems and provide the simulation details.

\subsection{ DMC algorithm for spherical particles}

 Colloids suspended in a solvent  undergo overdamped  Brownian motion with diffusive behavior,
for large enough times compared to the momentum relaxation time $\tau_M ^t$.
The latter quantity is set by the colloids mass $M$ and the translational friction coefficient $\gamma_t$ that depends on the particle size and shape,  its value being $3 \pi \eta \sigma $ for spherical objects of diameter $\sigma$ with stick boundary conditions: $\tau_M^t \equiv M/\gamma_t $. The resulting mean-squared displacement (MSD) of non-interacting colloids varies linearly with time  for $ t \gg \tau_M^t$, with a slope given by the  infinite-dilution diffusion coefficient $D_0^t= k_B T/ \gamma_t$. However, for interacting colloids in non-dilute suspensions,   different diffusion processes  should be distinguished, namely short-time $D_S^t$ and long-time $D_L^t$  diffusion. The distinction requires the introduction of the Brownian time-scale  $\tau_B$ defined as the time required for an isolated colloid to diffuse over its diameter $\sigma$, i.e.,  $\tau_B \equiv \sigma^2/ (6 D^t_0)$.  For relatively short times, larger than $\tau_M^t$  but smaller than the Brownian time-scale,   the colloids influence each others motions indirectly through the solvent flow field in which they move.   These solvent mediated hydrodynamic interactions may affect the short-time diffusion. If one ignores the hydrodynamic interactions, as in the subsequent analysis,
the short-time diffusion is  that of infinite-dilution diffusion coefficient $D_s^t=D_0^t$ \cite{Nagele,Felderhof}. On the contrary, the long-time diffusion $D_L^t$ that is defined  for $ t \gg \tau_B  $ is mainly determined by the direct interactions between colloids \cite{Medina,Nagele}.
For typical  colloids with diameters in the range 10 nm-1 $\mu$m, we have $\tau_B/\tau_M^t \gg 1$
with well separated diffusive regimes.

 Now, consider a MC procedure discarding hydrodynamic interactions, where each of $N$  interacting spherical particles in the simulation box  is shifted  by a random displacement chosen in the interval $[-\delta l, \delta l]$ along each Cartesian coordinate. The moves are accepted according to the Metropolis algorithm \cite{Frenkel}. Such a  simulation  mimics the Brownian motion of the colloids for time-scales $t \gg \tau_M^t$. One expects that the  mean-squared displacement (MSD) of a particle
 after $n$ cycles
$ \langle \Delta r^2(n) \rangle=1/N \sum_{i=1}^{N}  \langle |\vec{r}_i(n)-\vec{r}_i(0)|^2 \rangle $
varies linearly with the number $n$ of MC cycles, for both small values of $n$ corresponding
to the short time regime, and large $n$, albeit with a different slope.
The MSD   for sufficiently small $n$ is governed by the infinite-dilution diffusion  coefficient $D_0^t$, i.e., $\langle \Delta r^2(n) \rangle \approx 6D_0^t (n \delta t)$, provided that the amplitude of MC move $\delta l$ is chosen sufficiently small,  i.e., $\delta l \ll \sigma$ where $\delta t$ is the  physical time interval that each MC cycle corresponds to.  As a result, we impose that the relation between the MC clock and the real time  $\delta t$ can be obtained  from the slope of MSD in the small $n$ limit, i.e. the short-time diffusion of MC simulation.
 \begin{equation}\label{tscal1}
 \frac{\delta t}{\tau_B}=\lim_{n \rightarrow 0}\frac{ \langle \Delta r^2(n) \rangle}{n\sigma^2}
 \end{equation}
where  $\lim_{n \rightarrow 0}$ with $n$ an integer refers to the limiting behavior
of the MSD slope at  small $n$.
It has been noted that for  sufficiently small  $\delta l$,  $\langle \Delta r^2 (1)\rangle =A \delta l^2 $ where $A$ is the acceptance probability of the MC scheme \cite{Eduardo1}. It was therefore suggested that the time-scale corresponding to a MC cycle can be obtained as: $ \delta t= A  \delta l^2 /6 D_0^t  $ .  Scaled in terms of Brownian time, this equation can be written as
    \begin{equation} \label{tscal2}
\frac{\delta t}{\tau_B}= A \frac{ \delta l^2}{ \sigma^2}
   \end{equation}
which provides an alternative route against which our approach will be tested in section \ref{sec:assess}. This $A$-rescaling procedure will lead below to
the variant denoted $V_A$, while the diffusion-matching will be
referred to as variant $V_D$.
We will show that, quite expectedly, both methods become equivalent,  for sufficiently small  $\delta l$. However, enforcing
(\ref{tscal1}) instead of (\ref{tscal2}) allows us to extend the limit of applicability of  DMC towards larger values of $\delta l$.

\subsection{  DMC algorithm for anisotropic particles}

For anisotropic particles with orientational degrees of freedom and depending on the shape,
the diffusion in some directions is favored over some others, leading to the coupling of translational and rotational motions \cite {Lubensky}.  Henceforth, for
a meaningful dynamics, due account should be taken of
the anisotropy of  diffusion in the body-frame.  Similar to  the
translation-only case,
simulations based on  DMC   can produce the correct dynamics of the rotational degree of freedom for time scales larger than the damping time of angular velocity $\tau_M^r= I_r/\gamma_r$ where $I_r$ is the moment of inertia and $\gamma_r$ is the rotational friction coefficient. The exact value of $\tau_M^r$ depends on the size, shape of the particle and the axis of rotation under consideration, but its order of magnitude is the same as $\tau_M^t$. On the other hand, the time-scale for orientational relaxation $\tau_r$ is given by  the inverse of the infinite-dilution rotational diffusion coefficient $D_{0}^r=k_B T / \gamma_r$ leading to  $\tau_r/\tau_M^r=k_B T/I_r \gg 1$ for sufficiently large colloids.

\begin{figure}[ht]
\begin{center}
\includegraphics[scale=0.35]{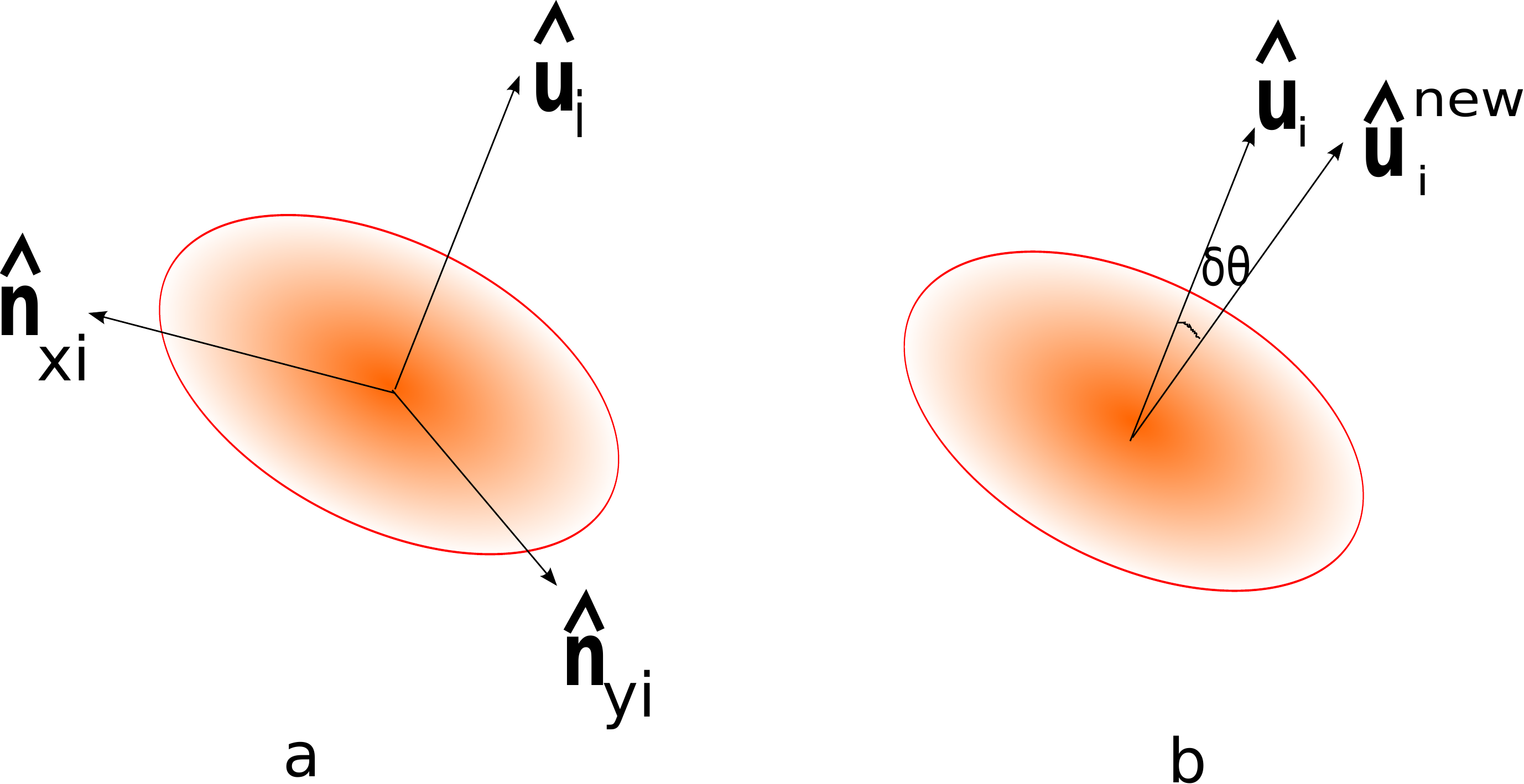}
\caption{  A schematic drawing  presenting a) an example of an axially symmetric object with its orientation vector $\widehat{u}_i$ together with its body frame defined by $\widehat{u}_i$ and two mutually perpendicular  vectors
$\widehat n_{x\,i}$ and $\widehat n_{y\,i}$. b)  rotation of orientation vector with an angle $\delta \theta$.}
\label{fig1}
\end{center}
\end{figure}

To illustrate the DMC implementation for anisotropic colloids, we focus here on axially symmetric particles, although the  generalization of our method to less symmetric objects
is straightforward.  We consider a system of  $N$ thin disks. Each of them can be identified by its center of mass position $\vec{r}_i$ and its  orientation unit vector $\widehat{u}_i$,  that is taken along the symmetry axis.   The translational (resp. rotational) diffusion tensor in the body frame is diagonal and consists of  one coefficient $D_{0||}^t$ (resp. $D_{0||}^r$)  for the direction parallel to $\widehat{u}_i$ and two identical coefficients $D_{0\bot}^t$ (resp. $D_{0\bot}^r$) for perpendicular directions.  In the following, we describe two variants of the same
MC algorithm, that  consists of simultaneous translational and rotational displacements  enforcing proper symmetry of the diffusion tensors. We ignore the rotations around the symmetry axis of the particle, as they cannot be detected in most of experiments; only rotations around axes perpendicular to symmetry axis are considered, i.e  rotations of the orientation  vector,  characterized by  the same coefficient $D_{0\bot}^r$  which for simplicity we denote as  $D_{0}^r$. The first variant, $V_A$, bears similarities with the
approach presented in \cite{Eduardo2} but differs in two respects:
first, no fine tuning of translational and orientational moves are required
here; second, we fully resolve the anisotropic dynamics at short times,
which is an essential prerequisite for the study of phases with orientational
order, such as nematic phases. The second variant is reminiscent of
the method put forward above for spherical particles.

We start by considering the limit of small MC increments,
for which the move acceptance probability is close to unity.
A general attempted translational move can be described by
$\delta \vec{r}_i=  \delta x \,\widehat{n}_{x\,i}+\delta y \,\widehat{n}_{y \,i}+
\delta z\, \widehat{u}_i $, where $\widehat{n}_{x \,i}$ and
$\widehat{n}_{y \,i}$ are two mutually perpendicular  arbitrary unit vectors that are normal to $\widehat{u}_i$ as well,  $\delta z $ is a random number in the interval $[-\delta l, \delta l ]$ and  $\delta x $, $\delta y$ are random numbers  in the interval  $[-\delta l', \delta l' ]$. For sufficiently small $\delta l$ and $\delta l'$,  the average MSD in the direction parallel to $\widehat{u}_i$ is   $\langle \delta z^2 \rangle \simeq \delta l^2/3$ and along perpendicular  directions  is  $ \langle \delta x^2 \rangle = \langle \delta y^2 \rangle \simeq \delta l'^2/3$. Imposing the symmetry of the diffusion tensor  provides us  with a relation between $\delta l' $ and  $\delta l$ , i .e.,
\begin{equation} \label{dl}
\frac{\delta l' }{\delta l} \approx \sqrt{\frac{D_{0\bot}^t}{D_{0||}^t}}
 \end{equation}
In addition, in the short-time diffusion regime, we should have
$\langle\delta z^2\rangle = 2 D_{0||} t$, so that the time increment corresponding to one single MC cycle reads
\begin{equation}
\label{dt}
\delta t = \frac{ \langle\delta z^2(1)\rangle}{2D^t_{0||}}
\end{equation}
If the orientation distribution for the ensemble of disks is
isotropic, a restrictive assumption,
we have $\langle \delta r^2 \rangle = 2 D_{0\bot} \delta l'^2
+ D_{0||} \delta l^2$.
Defining the average translational diffusion as $D_0^t=(2 D_{0\bot}^t+D_{0||}^t)/3$, the MSD can then
be simplified to $\langle \delta r^2 \rangle \approx \frac{D_{0}^t}{D_{0||}^t}\delta l^2$.

The change of orientation can be seen as  a random rotational displacement with an angle $\delta \theta$  in the interval $[0,\delta \alpha]$, such that $\widehat{u}_i^{new}. \widehat{u}_i= \cos \delta \theta$, see Fig. \ref{fig1} for an illustration. Such a rotation can be achieved as follows \cite{Daan-disks,Frenkel}: we generate a unit vector $\widehat{u}^{\prime}_i$  with an isotropic  random orientation  and obtain the new orientation vector as
\begin{equation} \label{MC-rot}
\widehat{u}_i^{new}= \mathcal{N}[(1-\delta \alpha) \widehat{u}_i+ \delta \alpha \widehat{u}^{\prime}_i]
 \end{equation}
where $\mathcal{N}$ ensures proper normalization. From this, one can calculate the correlation between the new and initial orientation vector in terms of $\delta \alpha$:
\begin{eqnarray} \label{MC-rot1}
\langle \widehat{u}_i^{new}.\widehat{u}_i \rangle = F(\delta \alpha) \equiv \frac{6+4(-3+\delta \alpha)\delta \alpha}{6(-1+\delta \alpha)^2} \quad \delta \alpha <0.5 \\ \nonumber
\simeq 1- \delta \alpha^2/3+ 2/3 \delta \alpha^3  \quad \delta \alpha  \ll 1
 \end{eqnarray}
where the average is taken over all the possible orientations of the random vector $\widehat{u}^{\prime}_i$	.
In addition, for the physical Brownian system under study,
we have $\langle \widehat{u}(t).\widehat{u} (0) \rangle= \exp(-2D_0^r t)$, \cite{Dhont}
so that we get from equation (\ref{MC-rot1}) that
\begin{equation} \label{MC-rot2}
\delta t\, =\, -\, \frac{\ln(F(\delta \alpha))}{2 D_0^r}
 \end{equation}
In the limit of small $\delta \alpha$,   the mean-squared angular displacement for diffusion of orientational vector  can be obtained from Eq. (\ref{MC-rot1}): $|\langle \widehat{u}(\delta \alpha)-\widehat{u} (0) \rangle|^2 \equiv \langle \delta \theta^2 \rangle \simeq 2 \delta \alpha^2/3$ (for $\delta \alpha \ll 1)$  that should be equal to $4 D_{0}^r A \delta t$.
The last step is to enforce consistency of time scales, equating  the  two relations for $\delta t$, i.e. Eq. (\ref{dt}) and Eq. (\ref{MC-rot2}). In doing so, we obtain
\begin{equation}
\delta t = \frac{ \delta l^2}{6D^t_{0||}}
\label{eq:tfict}
\end{equation}
and
the following constraint
between the amplitudes of translational and rotational moves:
\begin{equation}\label{amp}
\delta l = \sqrt{-3 \ln(F(\delta \alpha))}  \sqrt{\frac{D_{0}^r}{D^t_{0||}}}
  \end{equation}
In the limit of smaller $ \delta \alpha$, this simplifies into:
\begin{equation}\label{ratio}
\frac{\delta \alpha}{\delta l} \simeq \sqrt{\frac{D_{0}^r}{D^t_{0||}}}
  \end{equation}

We should now take due account of the fact that $A \neq 1$.
In the first variant of the approach, $V_A$, it is assumed that
the physical time increment is slowed down by rejected moves,
so that the generalized Eqs. (\ref{MC-rot2}) and (\ref{eq:tfict})
read:
\begin{eqnarray} \label{eq:diffA}
\delta t = A \frac{ \delta l^2}{6D^t_{0||}} \\
\delta t=-A \frac{\ln(F(\delta \alpha))}{2 D_0^r} \label{eq:autre}
\end{eqnarray}
Therefore, once the amplitude of  orientational moves $\delta \alpha$ has been chosen,
the amplitudes of translational moves  $\delta l$ and $\delta l'$
follow from Eq. (\ref{amp}) and Eq. (\ref{dl}). Physical time is given by (\ref{eq:diffA}) or equivalently (\ref{eq:autre}), where the acceptance
probability $A$ is computed on the fly in the simulation.
Alternatively, for variant $V_D$, we again impose (\ref{dl}) and (\ref{amp}),
but determine the physical time scale by imposing that
$\langle\delta z^2(1)\rangle$ computed in the simulation,
coincides with $2 D^t_{0||}t$ .

Before illustrating the applicability of our DMC algorithm, we provide
in the following section some details concerning the  systems simulated.

\subsection{Model systems and simulation details}
In the remainder, we investigate the dynamics of two model systems  by  means of DMC simulations:
hard sphere colloids (system A) and infinitely thin disks (system B) with diameters $\sigma$.
We take  into account the direct hard-core interactions by choosing at random a particle and generating a random trial MC move (including the rotational move for disks). We then reject the displacements that lead to an overlap with neighbors \cite{Daan-disks}.

The first system consists of $N$ particles  in a cubic box of length $L$, with  periodic boundary
 conditions. We took $N=1024$ and the  simulations were performed for  volume fractions
 $\Phi \equiv \pi \sigma ^3/ (6 N L^3)$ in
 the range $0.05-0.5 $.  The starting configuration was
 that of a  BCC crystal melted by an equilibration  run  of $2 \times 10^5$ MC cycles (one trial
 move per particle). The production runs for calculating the  mean-squared displacements
 consisted of  $1-5 \times 10^6$ cycles, depending on the volume fraction  and the amplitude
 of translational displacements $0.01 \le \delta  l/ \sigma \le 0.1$.

The second system consists of 500 disks again in a cubic simulation box with periodic boundary conditions.
For each reduced density $\rho^*=N \sigma^3/L^3$, a first equilibration run of $5-10 \times 10^4$  MC cycles starting with an initial configuration of disks on an FCC crystal with parallel orientations. The production runs for calculating the translational mean-squared displacements and orientational correlations were in the range of $10^6$ to $10^7$ cycles, depending on the density.
The  displacements amplitudes were in the range $2 \times 10^{-4} \le \delta l/ \sigma \le 0.2$. and
$ 0.005 \le \delta \alpha \le 0.324$.
The infinite-dilution  translational and rotational diffusion coefficients of disks  used in the simulations are
\begin{equation}
\label{eq:Dt}
D_{0||}=\frac{k_BT}{8 \eta \sigma} \quad \hbox{and} \quad
D_{0\bot}=\frac{3k_BT}{16 \eta \sigma}
\end{equation}
giving an average diffusion coefficient of $D_{0}^t=\frac{k_BT}{6 \eta \sigma}$.
On the other hand, we have for the rotational diffusion
\begin{equation} \label{eq:Dr}
D_{0r}=\frac{3kT}{4\eta \sigma^{3}}.
\end{equation}
These results are obtained from the general formula of diffusion coefficients of oblate spheroids \cite{Perrin,Pecora}  in the limit of vanishing length of semi-minor axis. Having described the methodology and simulation details, we
present below the results of our DMC simulations.

\section{ Assessment of the Dynamic Monte Carlo scheme }
\label{sec:assess}

In this section, we present  our DMC results for hard sphere self-diffusion, and then turn to thin disks. We compare and discuss the two different procedures for  mapping  MC time, i.e.  rescaling with acceptance probability or directly matching short-time dynamics with infinite-dilution diffusion tensor.

\subsection{ Dynamics of spherical colloids}
\begin{figure}[h!]
\begin{center}
\includegraphics[scale=0.29]{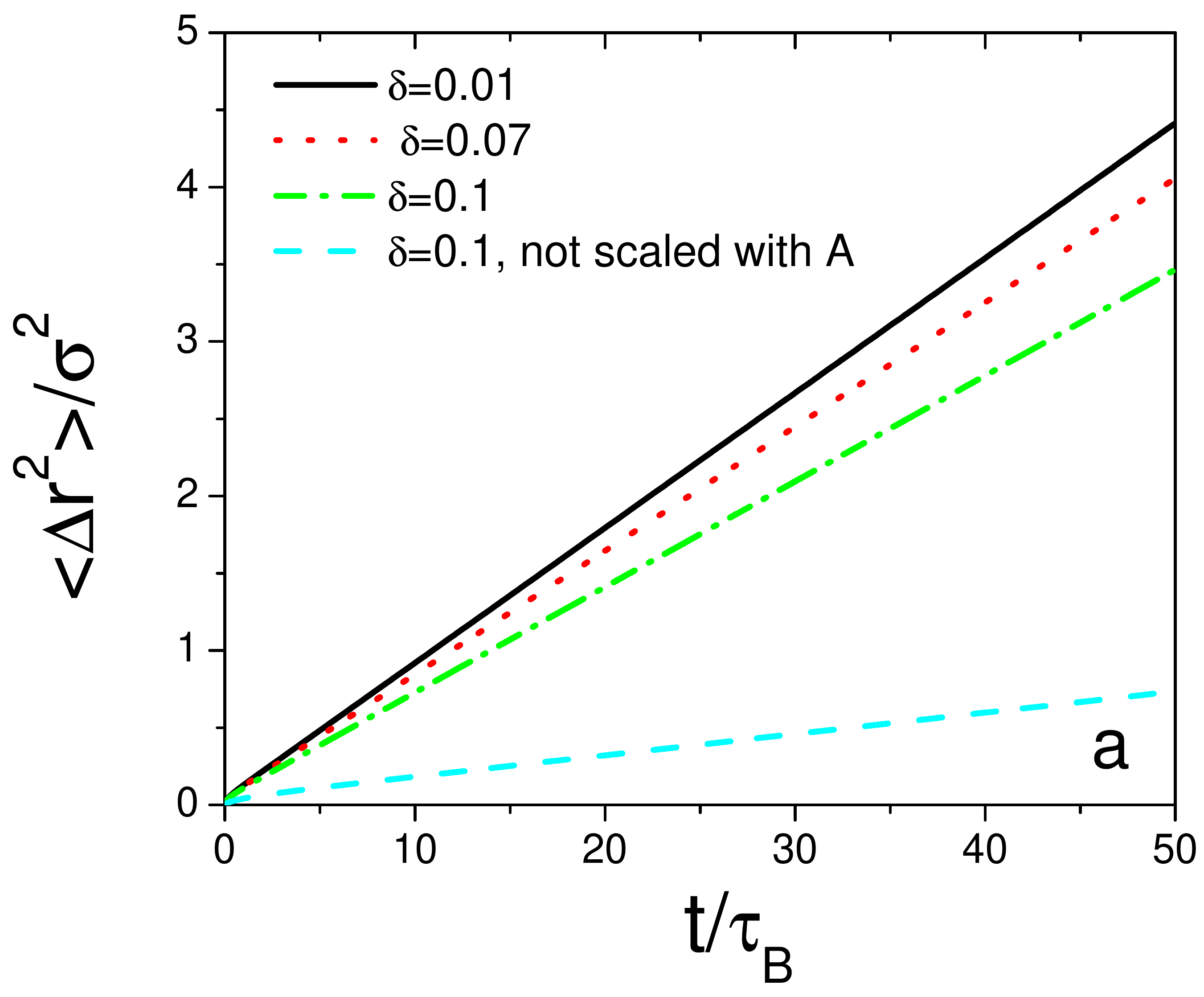}
\includegraphics[scale=0.29]{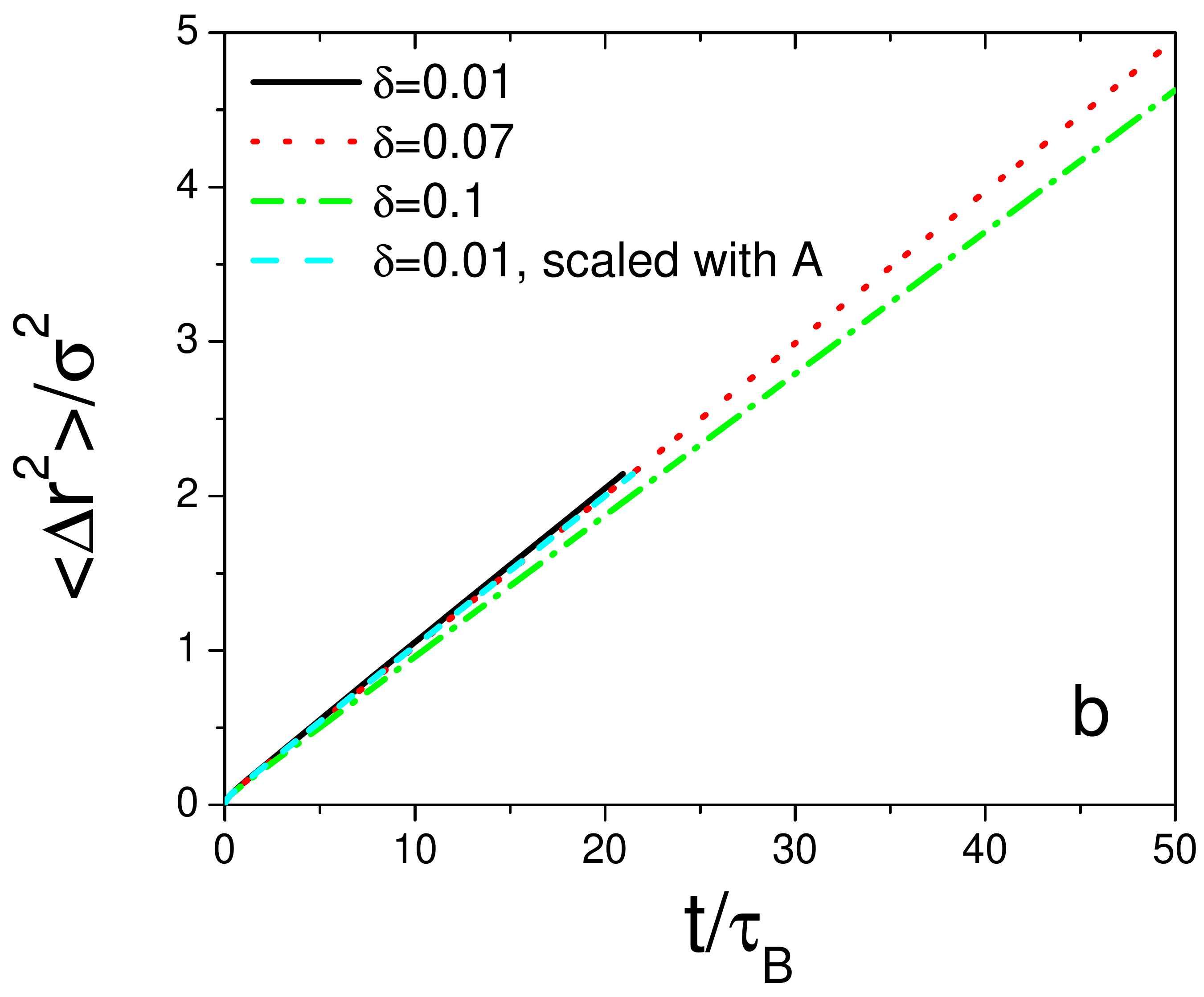}
\caption{   Mean-squared displacement obtained from DMC simulations for  a hard sphere system of volume fraction $\Phi=0.5$, as a function of time. The curves  correspond to three
different values of displacement amplitude   $\delta= \delta l /\sigma=0.01, 0.07$ and 0.1. The corresponding acceptance probabilities $A$ are: 0.84,  0.317 and 0.202, respectively. a)  $t/\tau_B$ is obtained by scaling with acceptance probability $A$ using Eq. (\ref{tscal2}). The lower dashed
curve shows the MSD data  for $\delta=0.1$ without $A$-rescaling. b) $t/\tau_B$ is obtained  from Eq. (\ref{tscal1}) by matching the MC short-time diffusion coefficient  to $D_0^t$. For comparison, the curve with $\delta=0.01$ and
$A$-rescaling [i.e. that shown in panel a)] is also plotted.
}
\label{fig2}
\end{center}
\end{figure}

We start by discussing  the  time-scale matching
in  a colloidal suspension of hard spheres. Figure \ref{fig2} shows the
time dependence of the mean
square displacement at  a relatively high volume fraction $\Phi=0.5$, for different values of MC displacement amplitude $\delta=\delta l/\sigma$. In  Fig. \ref{fig2}a, the physical time is obtained from $\delta ^2$ scaled  with acceptance probability, i.e.,  $t/\tau_B= n A \delta ^2$ as suggested in Ref. \cite{Eduardo1}.
As reported in \cite{Eduardo1}, such a procedure leads to a decent data collapse, the goal
being to obtain results that do not depend on $\delta$. In this respect, the collapse is only
partial, see e.g. the $\delta=0.1$ data that do not completely superimpose to those
for $\delta=0.01$.

In Fig. \ref{fig2}b, we have  obtained the physical time from the alternative method  leading to
Eq. (\ref{tscal1}). The graph shows that equating the short-time diffusion from MC with $D_0^t$ directly, allows a better collapse of MSDs for larger values of $\delta $ (where acceptance probabilities are smaller). We emphasize that for the small $\delta=0.01$, the two approaches yield the same results. However,  with  the present proposal, we can employ the DMC algorithm for relatively larger values of the increment $\delta$.

\begin{figure}[h!]
\begin{center}
\includegraphics[scale=0.29]{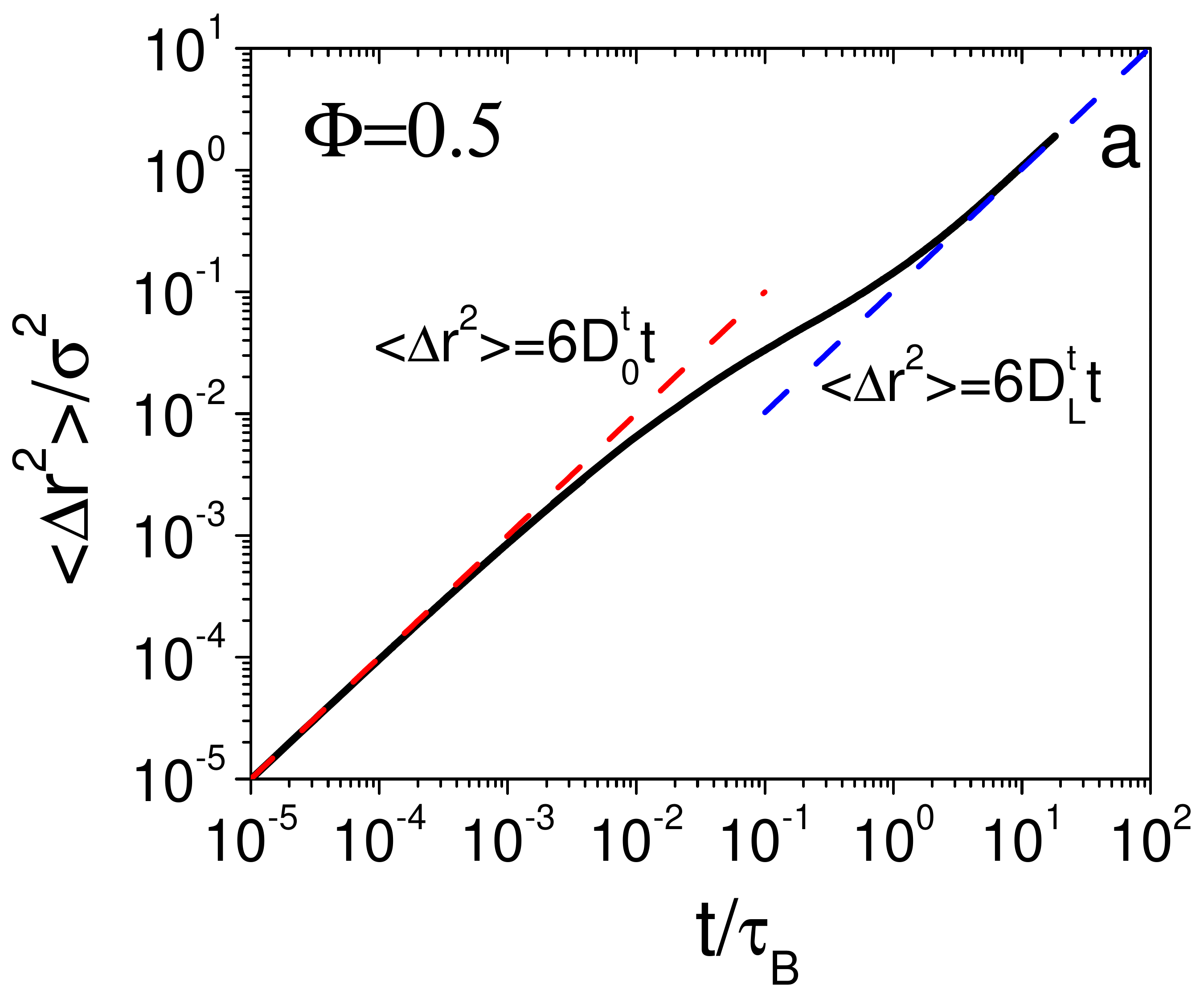}
\includegraphics[scale=0.3]{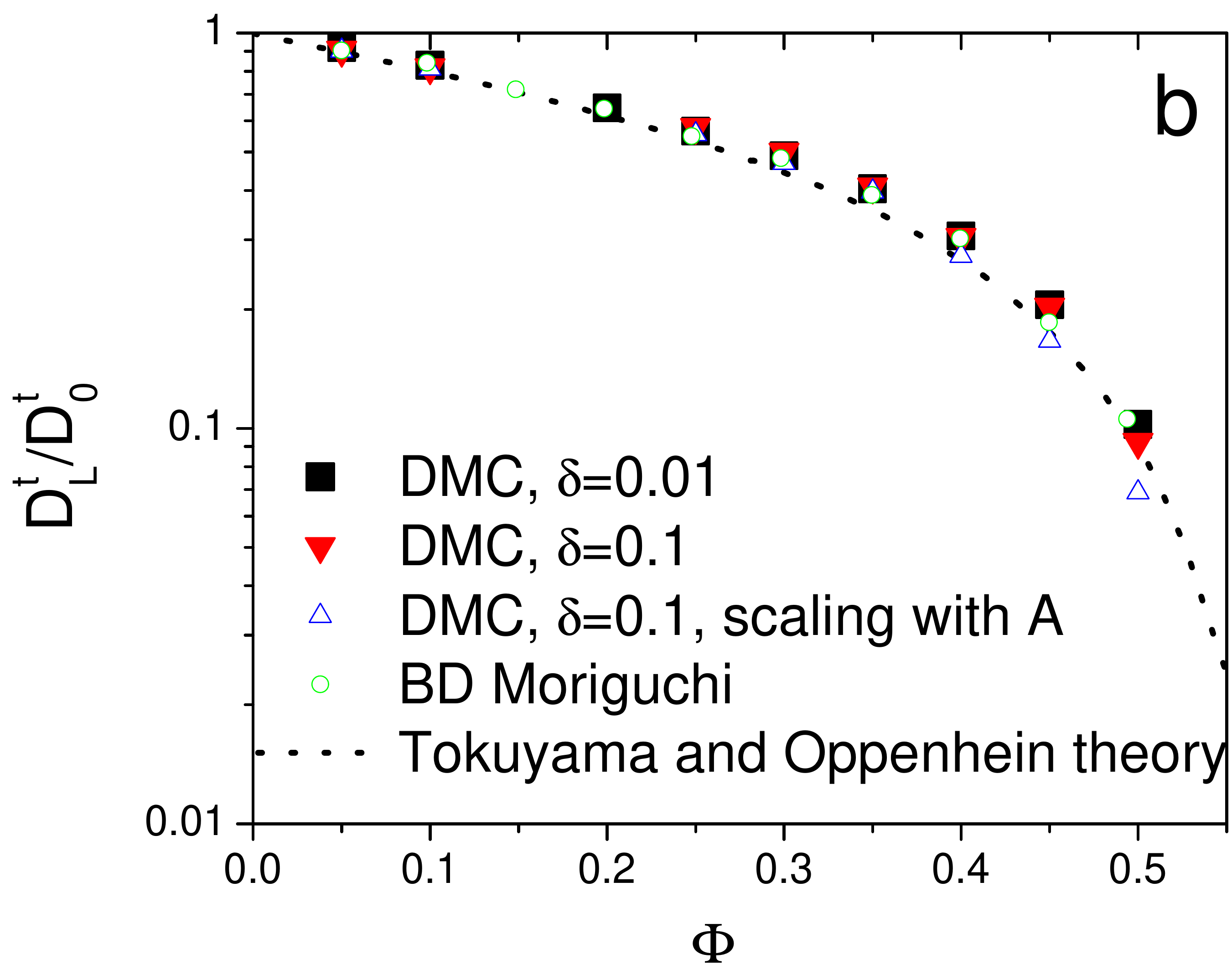}
\caption{  a) The mean square displacement obtained for $\Phi=0.5$ and $\delta=0.002$ clearly demonstrates a short-time slope of $6D_0^t$  (unity in scaled units) while at long time, the MSD grows with a reduced slope given by $D_L^t$. b) Long-time diffusion coefficient of hard sphere colloids without hydrodynamic interactions, obtained from DMC simulations with two methods, matching of short-time diffusion from MC with $D_{0}^t$ (solid squares  $\delta=0.01$ and triangle, $\delta=0.1$) and  with $A$-rescaling (open triangles, $\delta=0.1$). Also shown are hard sphere BD results from reference \cite{HSBD} (open circles). The short-dashed curve is for the Tokuyama and Oppenheim   formula \cite{Tokuyama}, that is used in  reference \cite{HSBD} to obtain the ratio of long to short time diffusion. }
\label{fig3}
\end{center}
\end{figure}

As discussed in the methodology section, the rationale behind our method is the fact that at short-times  ($t \ll \tau_B$), the particles diffuse freely  with diffusion coefficient $D_0^t$, while for long enough times ($t\gg \tau_B$), the MSD crosses over from free diffusion to a  slowed-down motion characterized by long-time diffusion $D_L^t$. This is illustrated in Fig. \ref{fig3}a,
where the reduction of diffusion coefficients is ten-fold.
To further test the reliability of our method, we have plotted in Fig. \ref{fig3}b
the long-time diffusion coefficient, extracted from  the slope of the MSD curve at long times, as a function of volume fraction $\Phi$ for different sampling
amplitudes. For comparison, we have also included the results obtained from scaling with $A$ for $\delta=0.1$ and Brownian dynamics results taken from reference \cite{HSBD}. The simulation data from event-driven BD  \cite{ED}  are fully consistent
with our DMC results (not shown).
While for $\delta=0.01$ the two approaches are equivalent (as illustrated in Fig. \ref{fig2}b),
some discrepancy is visible for $\delta=0.1$ and $\Phi> 0.35$. We conclude here that the
$A$-rescaling fares somewhat worse.
For completeness, we also have displayed the theoretical results of Tokuyama and Oppenheim
for the ratio of  long-time to short-time diffusion $D_{L}^t/D_{S}^t$ obtained for spherical particles when hydrodynamic interactions are accounted for.
Good agreement is found with the present simulation data that discard such
interactions, consistently with the findings of Ref. \cite{Medina}.

\subsection{\label{sec:level2b}  Dynamics of thin  colloidal disks  }

\begin{figure}[h!]
\begin{center}
\includegraphics[scale=0.29]{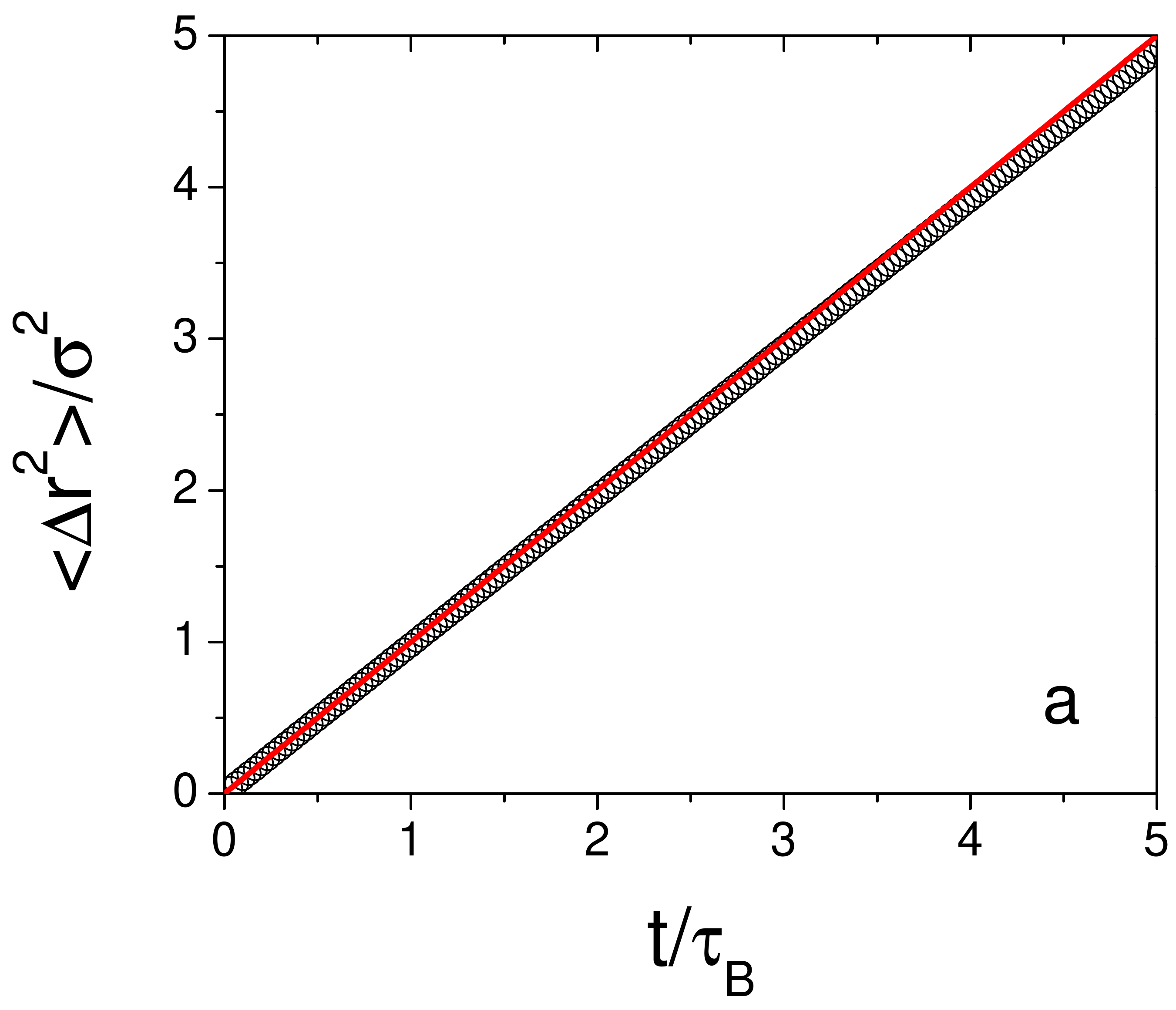}
\includegraphics[scale=0.29]{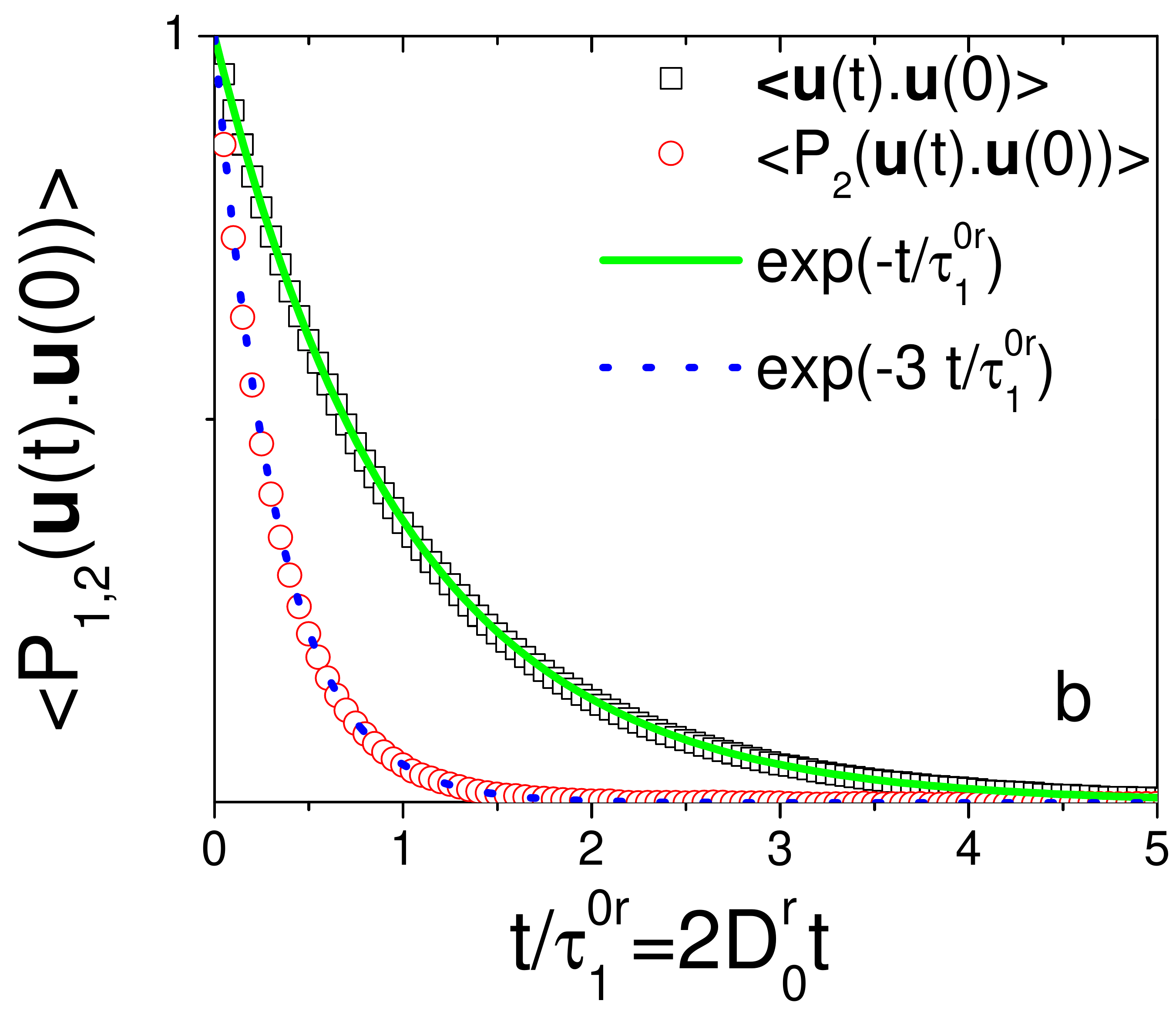}
\caption{ DMC simulations results of thin platelets at a low density  of $\rho^{*}=0.1$,  obtained for a  displacement amplitude of  $\delta= \delta l /\sigma=0.005$, for which the acceptance probability
is 0.9988. a) The MSD as a function of  $t/\tau_B= n  \delta l^2$ agrees well with the line of slope unity b)The first and second order time orientational correlations versus $t/\tau_{1}^{0r}=2 n D_{0}^r \delta \alpha^2/3 $  show a good agreement with $\exp(-t/\tau_{l}^{0r})$. }
\label{fig4}
\end{center}
\end{figure}

We now turn to the DMC simulations of thin disks.
First, we investigate the time behavior of  MSD and  orientational correlations for a low density system, and compare these results with the theoretical expectations for a freely diffusive particle.
As can be seen in Fig. \ref{fig4}a, the particles diffuse, as they should,
with the same diffusion coefficient at short and long times.
To quantify orientational dynamics, it is customary to define the
correlation functions $\langle P_l(\widehat{u}_i (t)\cdot \widehat{u}_i(0) ) \rangle$, where $P_l$ is the $l-$th order Legendre polynomial.
For a colloid in diluted conditions, these  orientational time correlation functions   decay exponentially with a relaxation time $\tau_l^{0r}=1/(D_0^r l(l+1))$  \cite{Pecora,Dhont}. Of particular interest among the correlation functions are those associated to $P_1$ and $P_2$, that are related to the dielectric properties of polar liquids and to the scattering of depolarized light, respectively \cite{Pecora}.
In panel (b) of Fig. \ref{fig4}, we have plotted both the first and the second order orientational correlation functions versus time. These
functions show a very good agreement with their analytical infinite
dilution counterparts. It should be noted that with the parameters
chosen in Fig. \ref{fig4} where the acceptance ratio is close to
unity, variants $V_A$ and $V_D$ coincide.

\begin{figure}[h!]
\begin{center}
\includegraphics[scale=0.29]{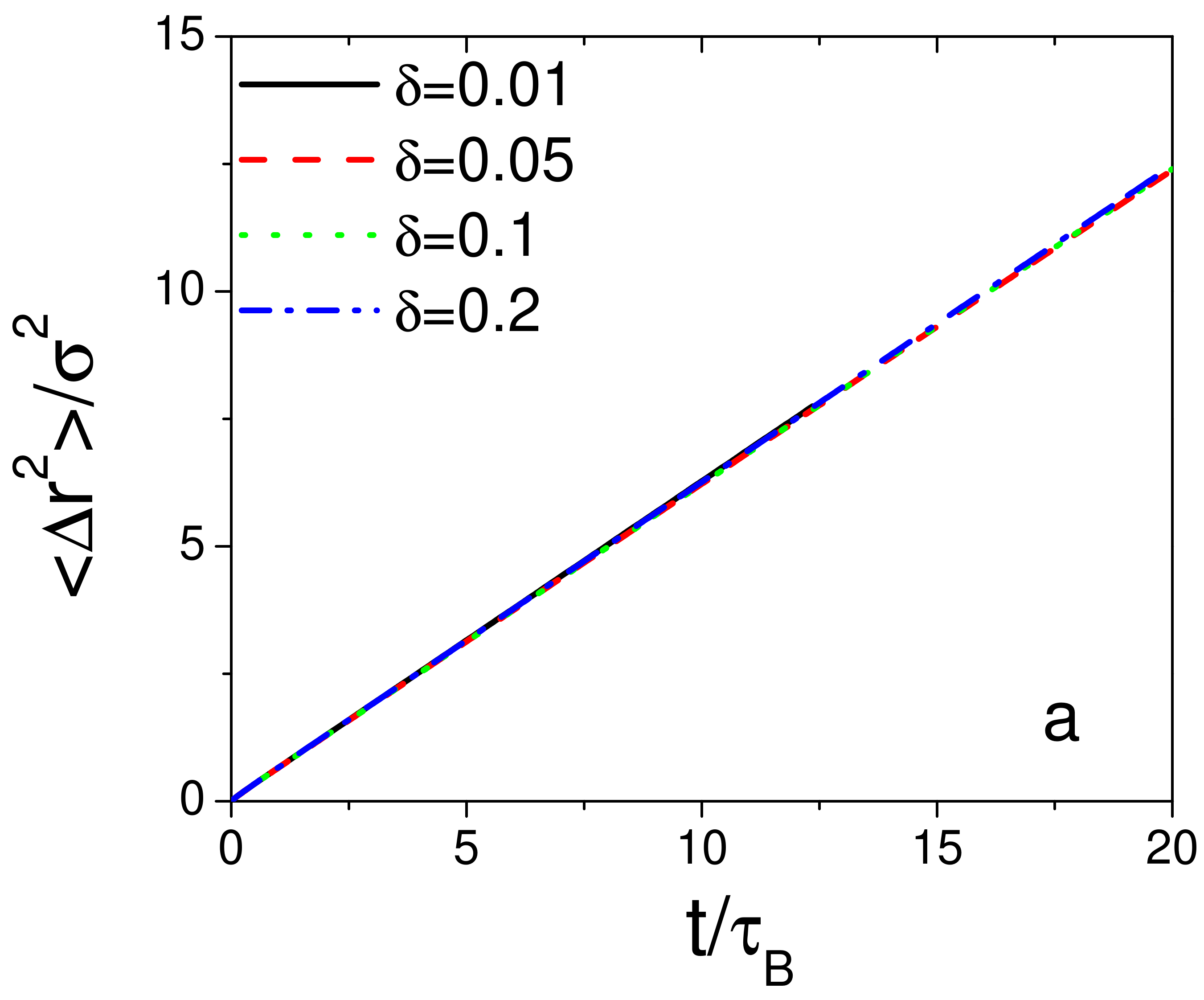}
\includegraphics[scale=0.29]{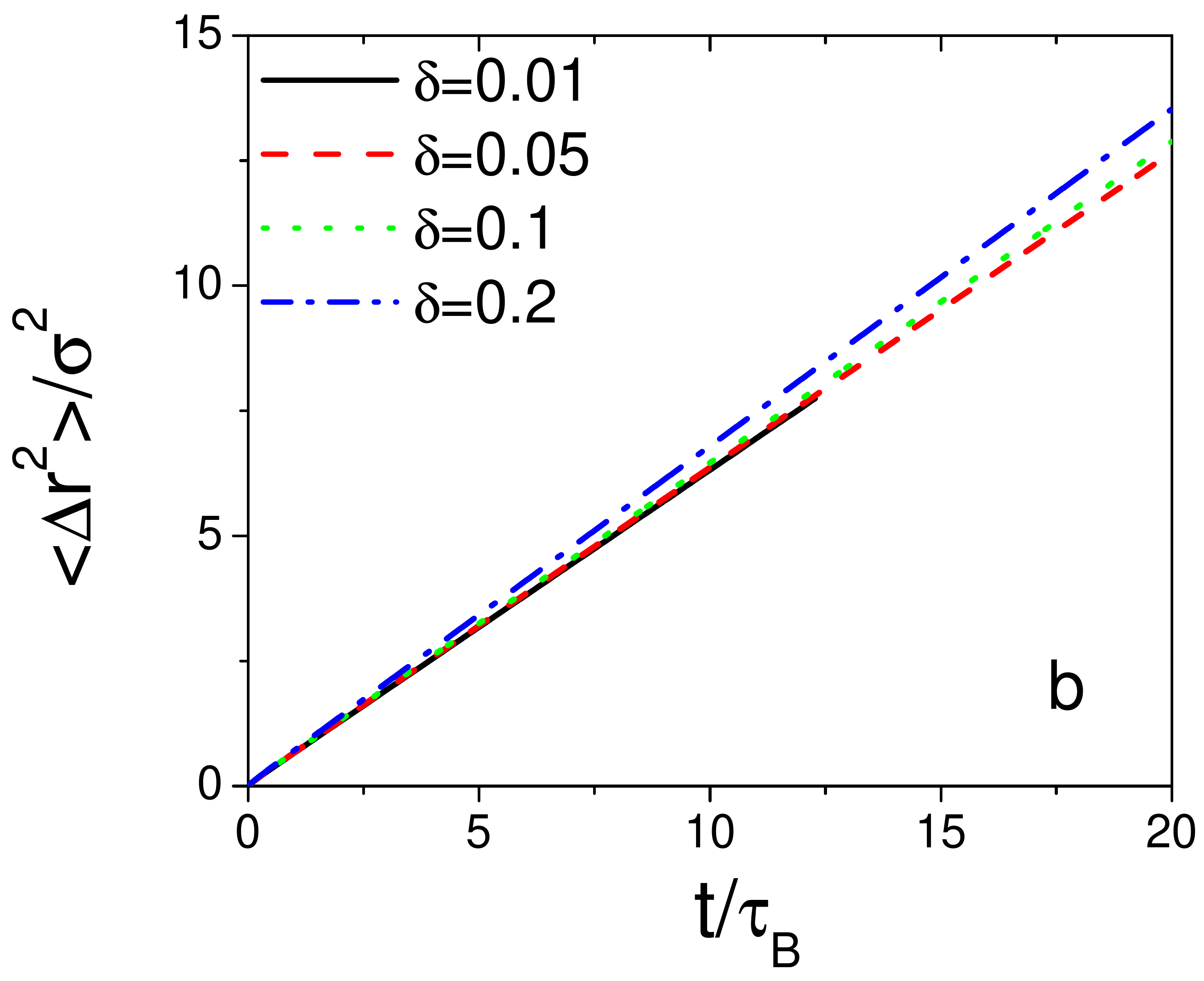}
\caption{Mean-squared displacement obtained from DMC simulations for  infinitely thin platelets of reduced density $\rho^{*}=2$, as a function of physical time. Different values of  sampling amplitudes were used:
 $\delta= \delta l /\sigma=0.01,0.05,0.1$ and 0.2 corresponding to $\delta \alpha=0.0225, 0.109,0.196$ and 0.324, respectively, and to the following
acceptance probabilities : 0.93, 0.68 and 0.48. a) $t/\tau_B$
is given by variant $V_A$  b) Results of variant $V_D$.
   }
\label{fig5}
\end{center}
\end{figure}

\begin{figure}[h!]
\begin{center}
\includegraphics[scale=0.29]{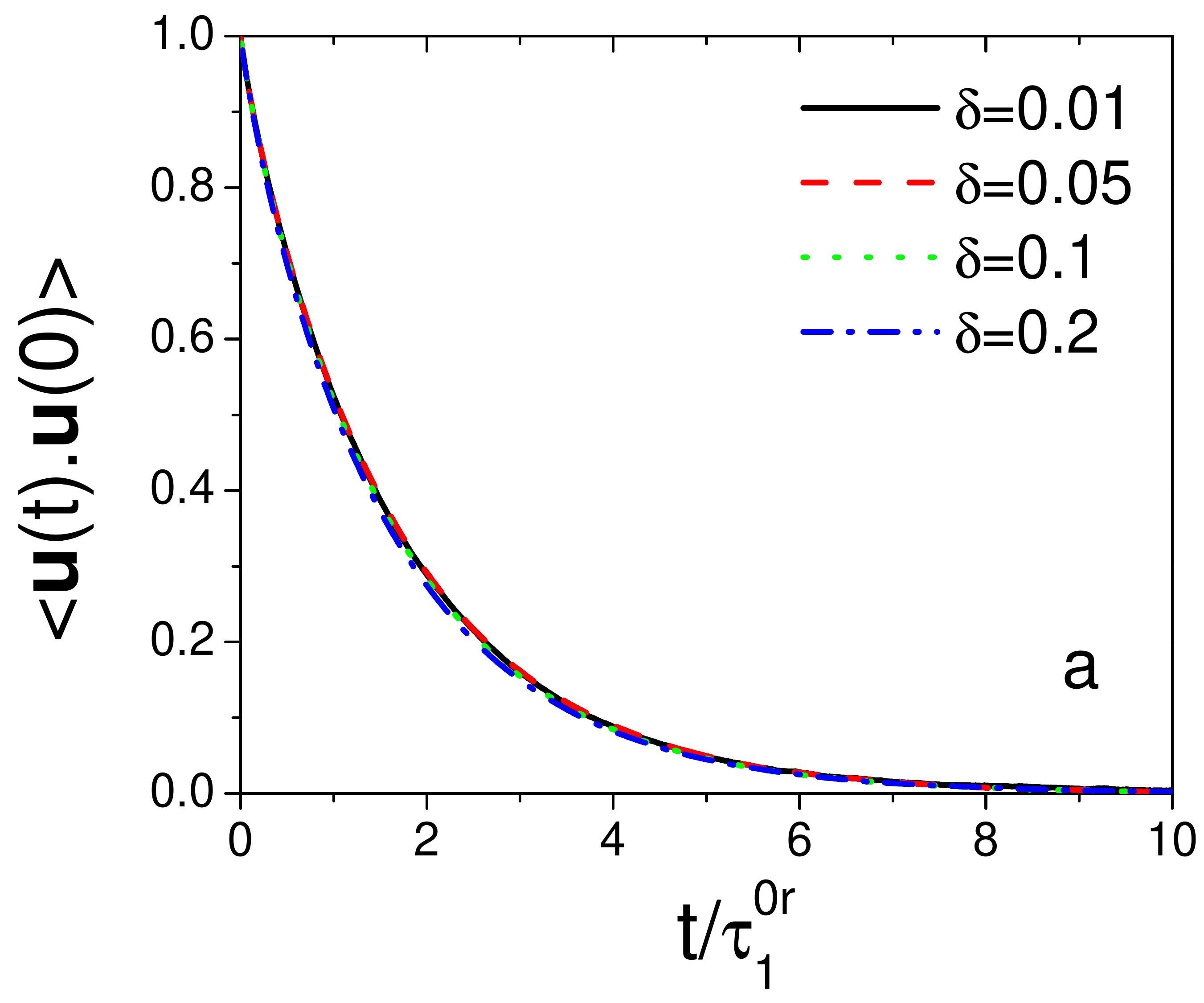}
\includegraphics[scale=0.29]{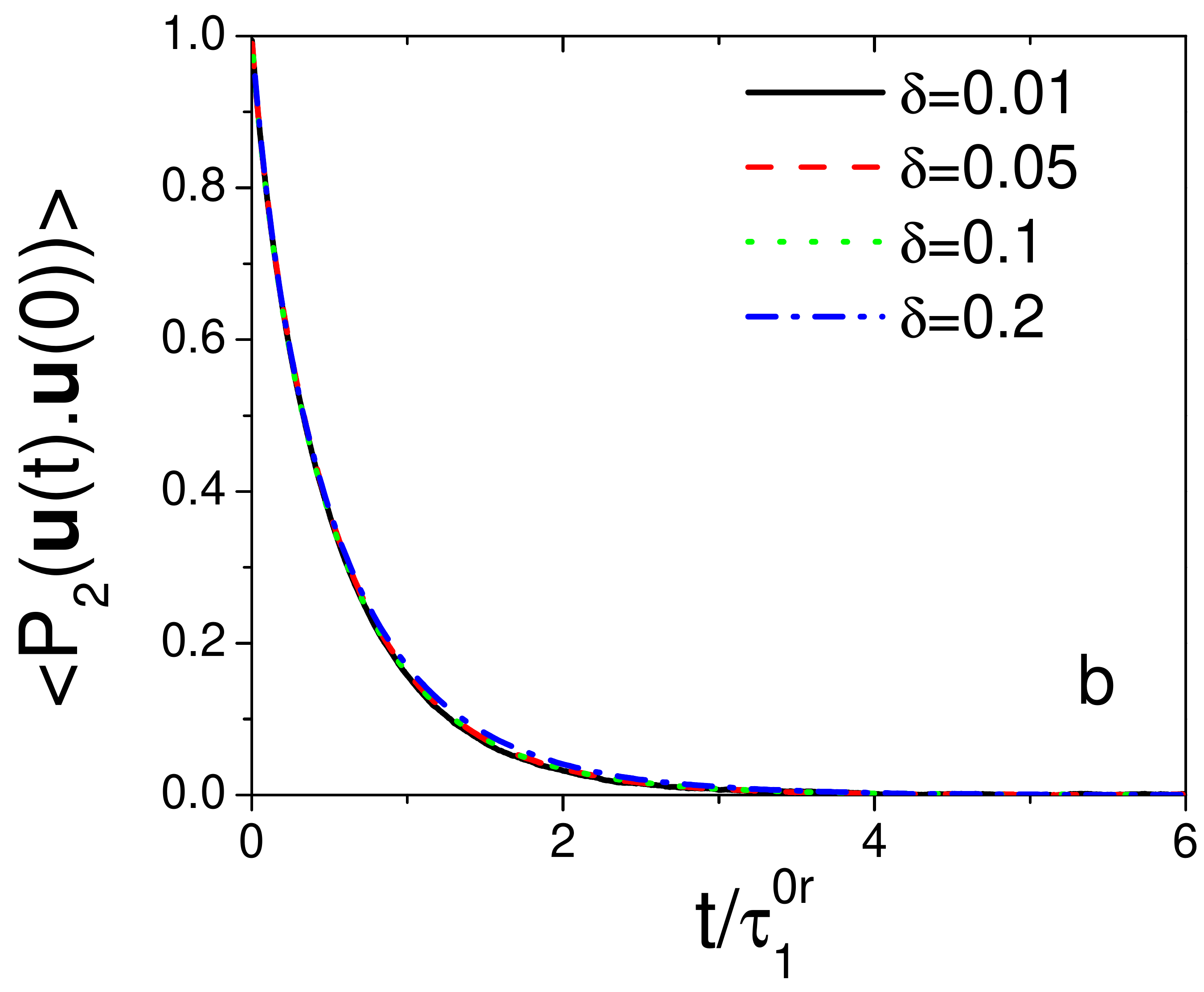}
\caption{  Same as Fig. \ref{fig5}, for orientational correlations.
In both cases, the results are reasonably independent of the
sampling parameters.}
\label{fig6}
\end{center}
\end{figure}

The next step is to explore the self-consistency of our two variants,
where the time behaviour generated should be independent of
the auxiliary parameters chosen for MC sampling. We have one such parameter,
say $\delta \alpha$, from which the other relevant increments
$\delta l$ and $\delta l'$ follow, see Eqs.
(\ref{amp}) and (\ref{dl}).
We first analyze   the behavior of  the translational self-diffusion. In figure \ref{fig5}, we have plotted  the MSD of  disks at  a density  $\rho^{*}=2$,
that is below the density of the isotropic-nematic transition
$\rho^{*}_{IN}=4$  \cite{Daan-disks}, as a function of  physical time  $t/ \tau_B$,  for different MC  sampling amplitudes. As discussed in
section \ref{sec:method}, we perform simultaneous translational and rotational moves. It appears that both variants $V_A$ [with results shown
in panel a)] and $V_D$ [results in panel b)] are satisfactorily
self-consistent, with a proper collapse of data. Relatively
large values of the sampling parameter are therefore acceptable,
and provide results of a comparable accuracy as more demanding
simulation with finer resolution. Upon closer inspection,
it can be seen that variant $V_A$ shows a somewhat smaller dispersion
of results that $V_D$. The analysis of orientational time correlations
corroborates this conclusion, see Fig. \ref{fig6}.
These conclusive tests allow us to study the density dependence
of long time diffusion in a system of disks, and in particular
the effect of a phase transition crossing.


\section{Anomalous diffusion in the nematic phase of thin disks}
\label{sec:anomalous}

\begin{figure}[h!]
\begin{center}
\includegraphics[scale=0.29]{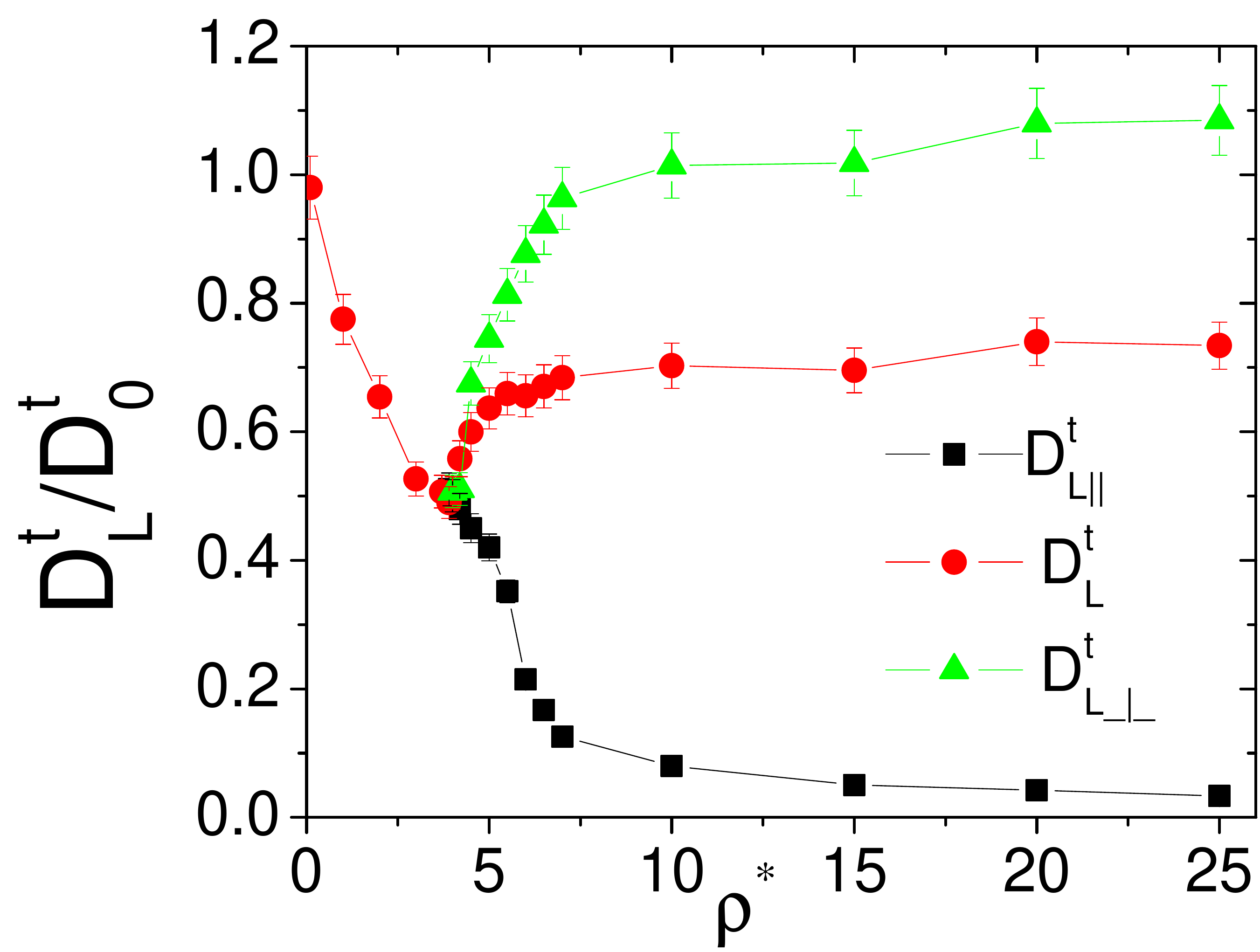}
\includegraphics[scale=0.29]{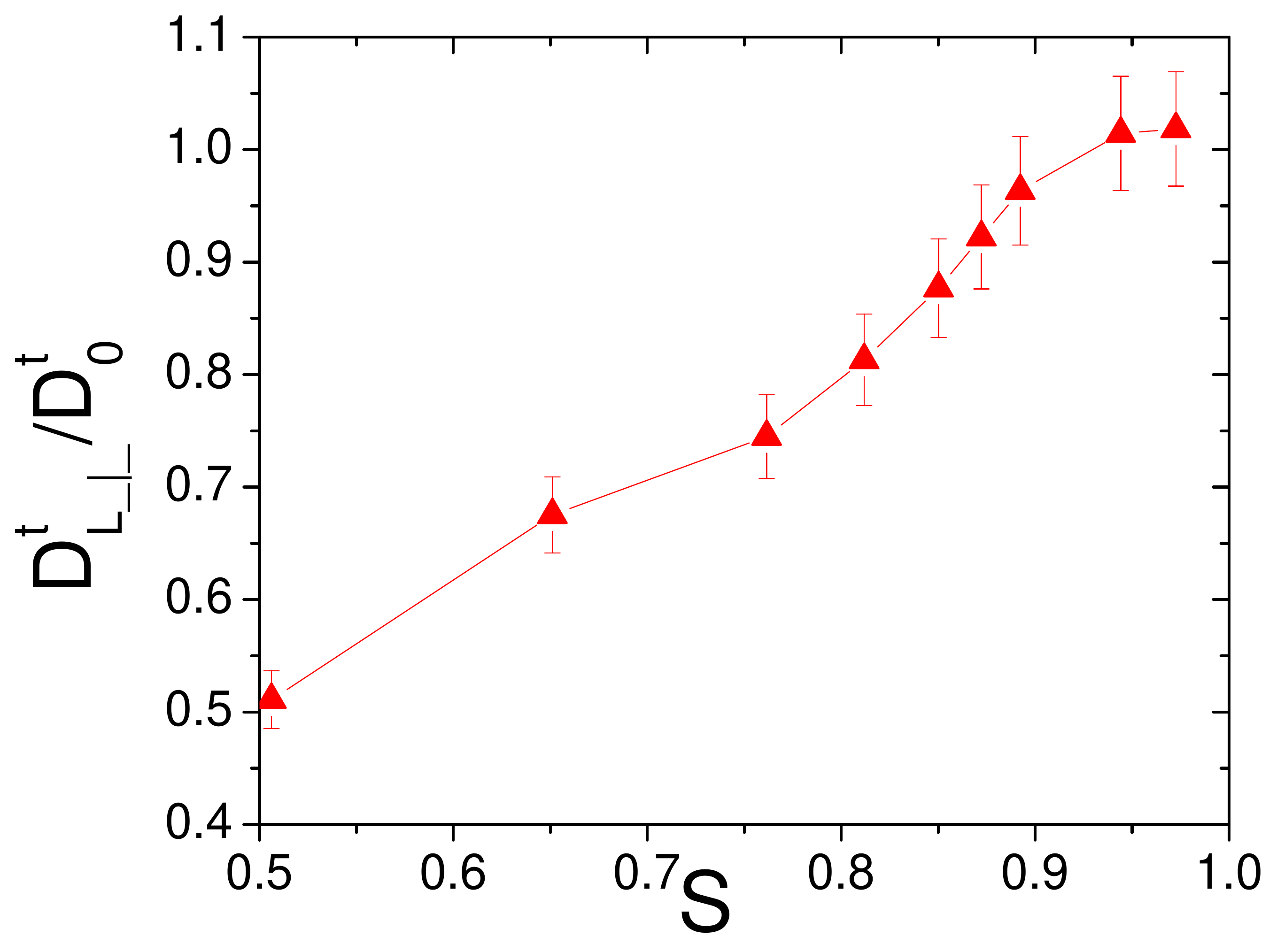}
\caption{a) Long-time average   diffusion coefficient  $D_L^t$  (in  both isotropic and nematic phases) and  long-time diffusion coefficients in directions parallel $D_{L||}^t$ and perpendicular $D_{L \bot}^t$ to the nematic director for  densities $\rho^*  \ge 4$. b) $D_{L \bot}^t$ as a function of nematic order parameter. }
\label{fig7}
\end{center}
\end{figure}

Thin platelets undergo an isotropic-nematic transition upon increasing the density \cite{Daan-disks}, and  it is interesting to  see  how the long-time translational  diffusion and orientational relaxation are affected. To this end, we have performed systematic DMC simulations with sufficiently small displacement amplitudes and have obtained both long-time translational self-diffusion coefficient and orientational relaxation time  as a function of density.
Fig. \ref{fig7}a depicts the long-time translation self-diffusion coefficient $D_L^t$.  We find that increasing the density, $D_L^t$ decreases up to the transition point. However, in the nematic phase
not only  the short-time diffusion  is anisotropic,  but also, the long-time diffusion becomes anisotropic with respect to the nematic director. Upon further increasing  the density, we observe that the diffusion coefficient in the direction perpendicular to nematic axis $D_{L\bot}^t$ grows while the  parallel component $D_{L||}^t$ decreases  significantly. As can be seen from Fig. \ref{fig7}a,  $D_{L \bot}^t$ approaches  the free diffusion coefficient of disks
$D_{0\bot}^t/D_{0}^t=9/8=1.125$  [see Eq. (\ref{eq:Dt})] in the limit of very high densities. In  Fig. \ref{fig7}b, we have plotted  $D_{L \bot}^t$ versus nematic order parameter $S$. As demonstrated by  this figure the more the disks become aligned, the larger is the perpendicular component of the diffusion in contrast to $D_{L||}^t$ that becomes very small:
topological constraints due to the excluded volume  constrain the disks to move in a  caging slab of parallel neighboring particles.
We also emphasize that for all results presented in this section,
variants $V_A$ and $V_D$ provide strictly identical
results.

\begin{figure}[h!]
\begin{center}
\includegraphics[scale=0.29]{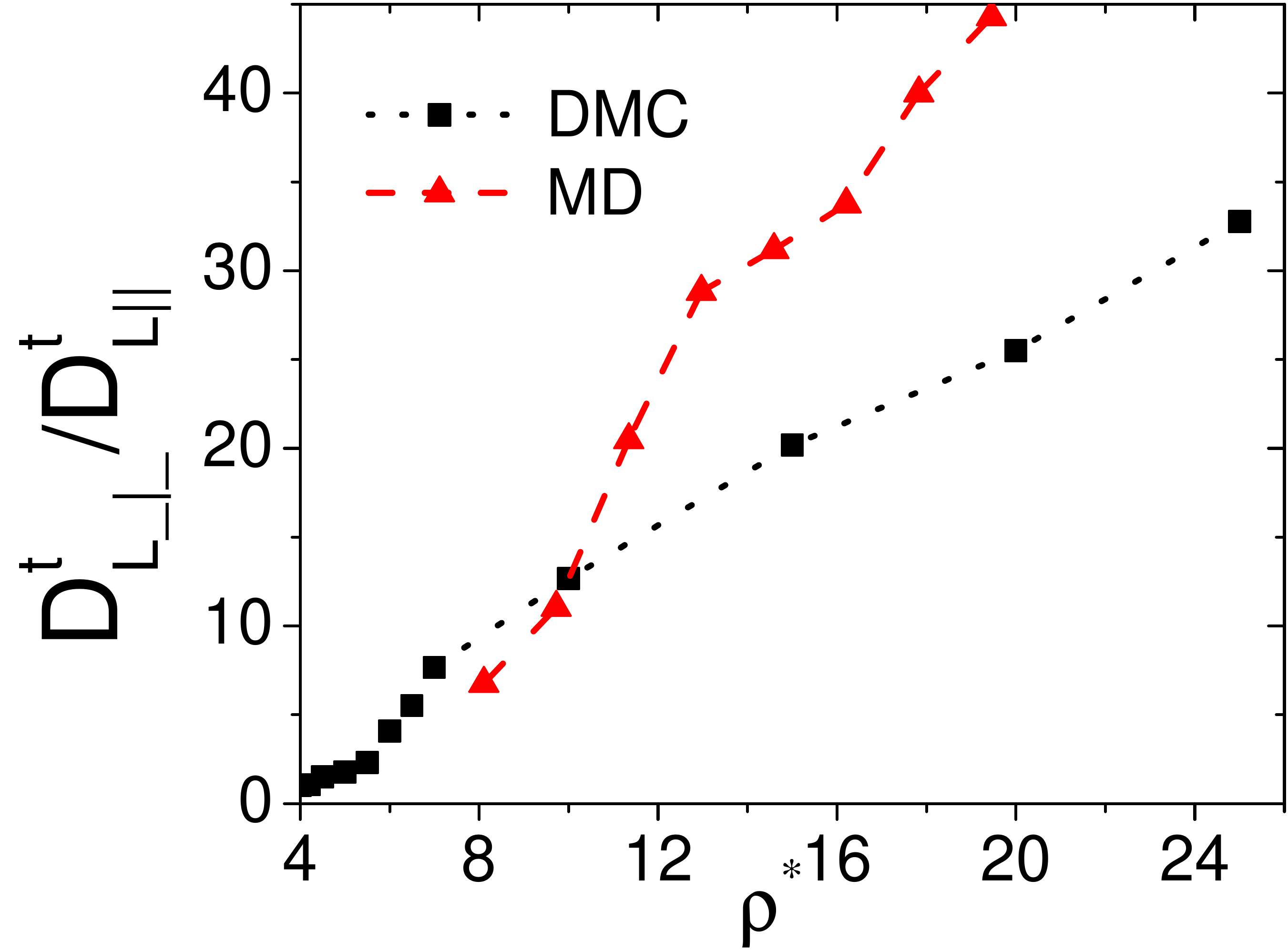}
\caption{Ratio $D_{L \bot}^t/ D_{L ||}^t$  for hard disks as a function of density. DMC data are compared with their MD counterpart of Ref. \cite{Daan-disks1}. }
\label{fig8}
\end{center}
\end{figure}

  At this point it is interesting to compare our DMC results  with molecular dynamics  (MD) simulations of thin disks  \cite{Daan-disks1}, where  an anisotropic diffusion in the nematic phase has also been observed \cite{Daan-disks1}. However, one should keep in mind that the model in
this work
is not  equivalent to ours.  In our DMC simulations, we mimic the presence of
an  underlying solvent through the stochastic nature of MC moves.
On the other hand, there is no solvent and hence the short-time
diffusion is replaced by a ballistic regime in the MD approach of
Ref. \cite{Daan-disks1}. For a quantitative comparison,
Fig. \ref{fig8} shows the ratio $D_{L \bot}^t/ D_{L ||}^t$ as a function of density. It seems that the two models have a different limiting behavior at high densities, where the MD data exhibit enhanced anisotropy.

\begin{figure}[h!]
\begin{center}
\includegraphics[scale=0.31]{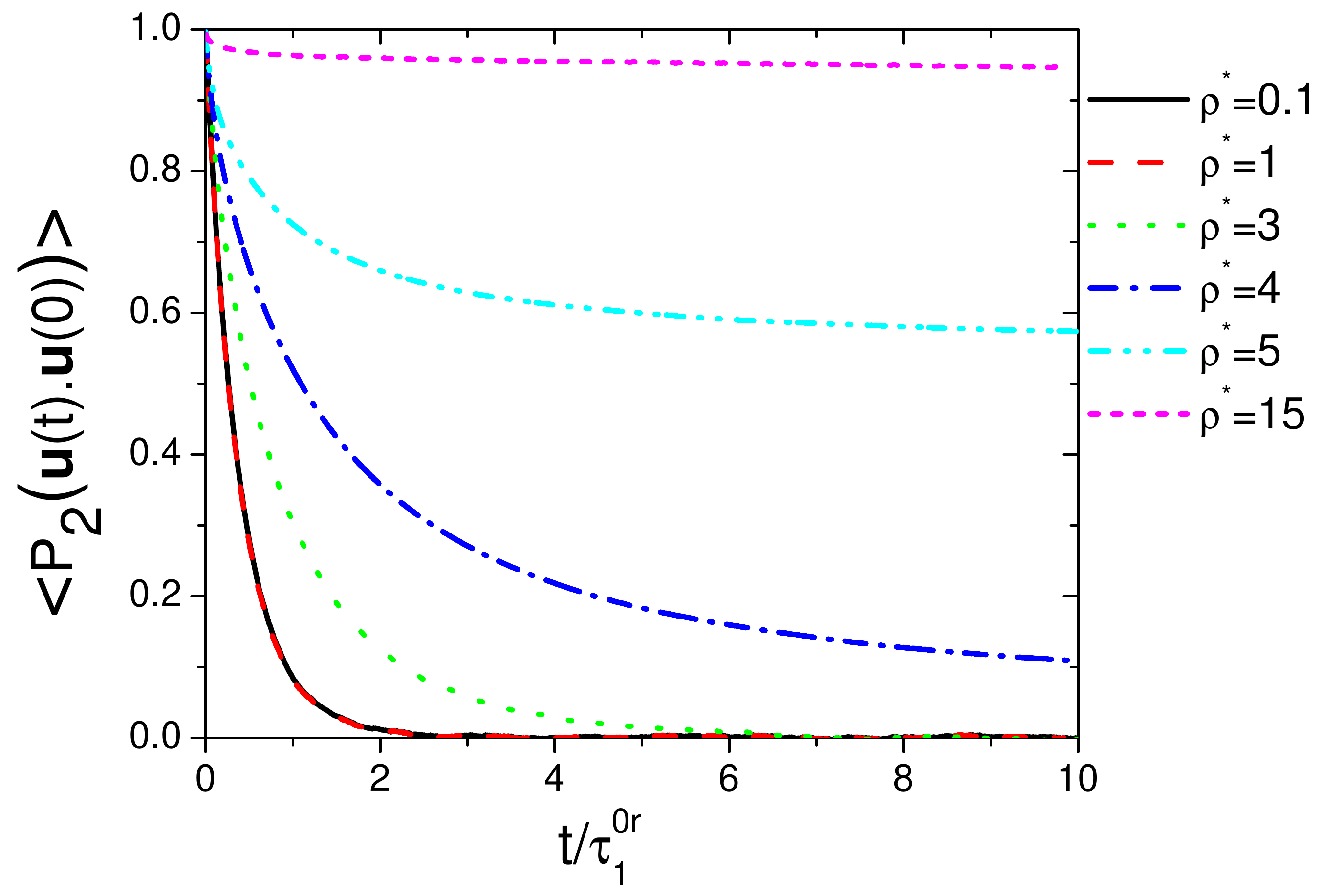}
\includegraphics[scale=0.27]{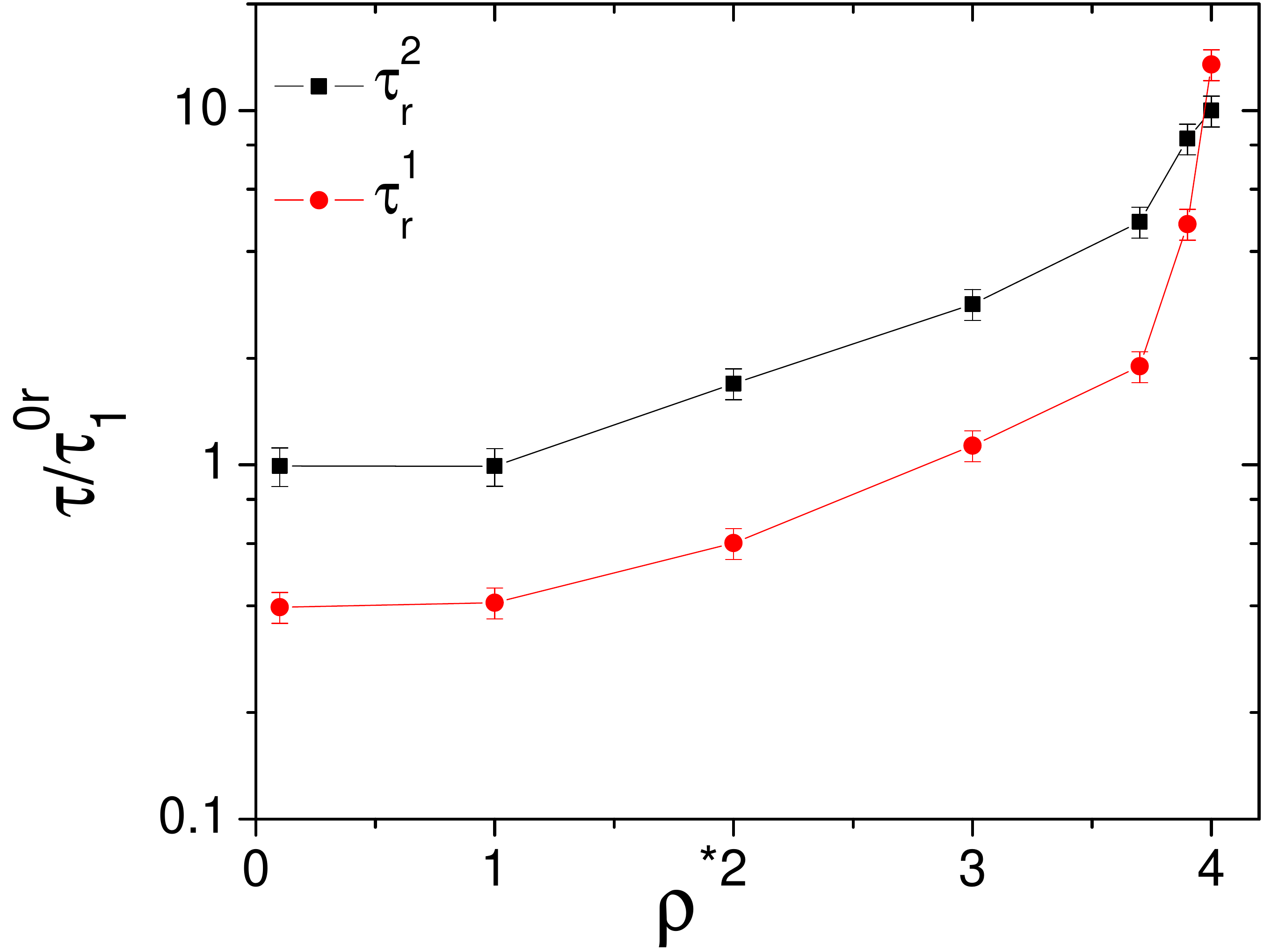}
\caption{ a) The second-order  orientational correlation function for different densities of disks b) The first  and second-order rotational relaxation times in the isotropic phase, as a function of density.
}
\label{fig9}
\end{center}
\end{figure}

We also have investigated the evolution of orientational time correlations with density, see Fig. \ref{fig9}a. We observe that the orientational time-correlations decay exponentially in the isotropic phase, while they  become non-ergodic in the nematic phase and develop a plateau at long times whose value is equal to the square of nematic order parameter $S$. To quantify the development of relaxation time with density, we have fitted the orientational correlation functions in the isotropic phase with an exponential, and obtained the corresponding relaxation time for the first and the second-order correlations as depicted in Fig. \ref{fig8}b. As expected, relaxation times grow with density  upon approaching the isotropic-nematic transition.

\section{Summary and Conclusions}
\label{sec:concl}

To summarize, we have presented  a dynamic Monte-Carlo algorithm (DMC) for both spherical and  anisotropic colloids. In each case, we have discussed the procedure for matching the Monte Carlo time scale
with its physical counterpart.
In the case of spherical particles, we found that matching the short-time diffusion from DMC to the infinite-dilution diffusion coefficient leads to a better convergence of results than acceptance rate based schemes,
for relatively large values of displacement amplitude.
A slightly better agreement of the long-time diffusion coefficient with  Brownian Dynamics data available in the literature was thereby achieved.

For anisotropic colloids, we presented two variants of the DMC algorithm that takes into account the anisotropy of short-time diffusion into account.
As for spheres, one is acceptance-rate based ($V_A$), and one relies on short-time diffusion matching ($V_D$). Both routes are new in their present formulation,
although variant $V_A$ shares common features with the approach
of Refs. \cite{Eduardo1,Eduardo2}. A key point is that the appropriate ratio
of translational and rotational move amplitudes is enforced, which
leads to the proper short-time diffusive behaviour.
We have tested the self-consistency of both variants, that give
similar results for a system of thin platelets in three dimensional space.
The method was finally employed to investigate the evolution of the
long-time diffusion coefficient and orientational correlation functions with density. The anisotropy of the long time translational diffusion tensor
was characterized in the nematic phase.
While diffusion along the nematic axis becomes small when nematic
ordering is more pronounced, it is enhanced in the perpendicular direction.

\begin{acknowledgments}
We wish to acknowledge the  support of Foundation Triangle de la Physiques and IEF Marie-Curie fellowship. We are also grateful to Patrick Davidson, Pierre Levitz and Jean-Jacques Weis for fruitful discussions.
\end{acknowledgments}

\bibliography{DMC}
\bigskip

\textbf{\large{List of symbols}}\\
\begin{itemize}
\item $ N $: number of particles in the simulation box of size $L$;
\item $ M $: Mass of particles;
\item $ \sigma=2R $: Sphere or disk diameter;
\item $ n $: number of Monte Carlo cycles, where a cycle is defined as one MC move per particle;
\item $ \Phi \equiv \pi \sigma^3 /(6 N L^3) $: volume fraction of spheres;
\item $ \rho \equiv N/L^3  $:  number density of thin disks;
\item $ \rho^*= \rho \sigma^3  $: dimensionless number density of thin disks;
\item $ \gamma_{t(r)}$:  translational  (rotational) friction coefficient;
\item $ \tau_M^{t} = M/ \gamma_t$: time-scale for which momenta of Brownian particles have relaxed;
\item $ \tau_M^{r} = I_r/ \gamma_r$: damping time of angular velocity for rotational Brownian particles;
\item $ D_0^t=k_BT/ \gamma_t $: infinite-dilution translational diffusion coefficient of spheres or average  translational  diffusion coefficient of anisotropic particles; for disks, we have
$D_0^t = 2 D_{0\bot}^t/3 + D_{0||}^t/3$.
\item $ D_S^t $: short-time translational diffusion coefficient;
\item $ D_L^t $: long-time translational diffusion coefficient;
\item $ \tau_B \equiv \sigma^2/ (6 D^t_0) $: Brownian time-scale, required for diffusing over a distance equal to the particle size;
\item $ D_{0\bot}^t=k_BT/ \gamma_t^{\bot} $: infinite-dilution translational diffusion of an axially symmetric particle in the direction perpendicular to the symmetry axis;
\item $ D_{0||}^t=k_BT/ \gamma^{||}_t $: infinite-dilution translational diffusion of an axially symmetric particle in the direction parallel to the symmetry axis;
\item $ D_{L \bot}^t $: Long-time translational self-diffusion of axially symmetric particles in the direction perpendicular to the nematic axis;
\item $ D_{||}^t $: Long-time translational self-diffusion of axially symmetric particles in the direction parallel to the nematic axis;
\item $ D_{0\bot}^r \equiv  D_{0}^r $: infinite-dilution rotational diffusion of axially symmetric particle in the direction perpendicular to the symmetry axis;
    $\tau_r=1/(2D_{0}^r)$:  time-scale for relaxation of orientation vector;
\item $ D_{0||}^r =k_BT/ \gamma^{||}_r=k_BT/ \gamma^{\bot}_r$: infinite-dilution rotational diffusion of axially symmetric particle in the direction parallel to the symmetry axis;
\item $\langle \Delta r^2  \rangle $: mean-square displacement (MSD);
\item $\langle \delta r^2  \rangle \equiv \langle \Delta r^2 (1) \rangle $: mean-square displacement after one MC step;
\item $\langle \delta \theta^2 \rangle   $: angular mean-square displacement;
\item $A$: acceptance probability;
\item $\delta l $:  amplitude of the translational displacement
\item $\delta= \delta l /\sigma $
\item $\delta \alpha $: maximal amplitude of the rotational  displacement
$\delta\theta$;
\item $\delta t $:   physical time interval corresponding to one MC cycle;
\item $ \tau_{l}^{r} $: the  relaxation time of the orientational time correlation  functions $\langle P_l(\widehat{u}(t) \cdot \widehat u(0) ) \rangle$;
\item $ \tau_{l}^{0r}=1/(l(l+1)D_{0}^r) $: the $l$-th order orientational relaxation time of an isolated particle;
\item $S$: nematic order parameter.
\end{itemize}
\end{document}